# Single Photon Adiabatic Wavelength Conversion

Stefan Preble[a)], Liang Cao, Ali Elshaari, Abdelsalam Aboketaf, Donald Adams

*Microsystems Engineering, Kate Gleason College of Engineering, Rochester Institute of Technology, Rochester, New York 14623, USA*

Adiabatic wavelength conversion is experimentally demonstrated at a single photon power-level using an integrated Silicon ring resonator. This approach allows conversion of a photon to arbitrary wavelengths with no energy or phase matching constraints. The conversion is inherently low-noise and efficient with greater than 10% conversion efficiencies for wavelength changes up to 0.5nm, more than twenty times the resonators line-width. The observed wavelength change and efficiency agrees well with theory and bright coherent light demonstrations. These results will enable integrated quantum optical wavelength conversion for application ranging from wavelength-multiplexed quantum networks to frequency bin entanglement.

The ability to shift the wavelength of light at the single photon level is critical for a number of quantum information applications. For example, it would enable channel control in wavelength division multiplexed quantum key distribution networks,[1,2] it could be used to realize a quantum analog of frequency-shift keying,[3] to control frequency-bin entanglement,[4] and would allow for the tuning of a photon's frequency to match the resonance of optical cavities, vacancy centers or quantum dots.[5,6] Single photon wavelength conversion has traditionally been realized using nonlinear frequency mixing in long crystals or optical fibers.[7–11] However, there has been considerable interest in realizing chip-scale based quantum information systems[12,13] to achieve robust integration and low cost. At the chip level there has been recent progress in realizing single photon wavelength conversion using four wave mixing in Silicon-compatible waveguides.[14] However, all approaches based on optical nonlinearities have the drawback of requiring high powered pump lasers, which will be challenging to filter away from the weak photon signals on a chip. Recently an alternative approach using cavity opto-mechanics to mediate the wavelength conversion process has been demonstrated.[15,16] This approach has the drawback of having to cool the cavity to low temperatures in order reduce thermal mechanical noise to single photon levels.

Here we propose and experimentally demonstrate adiabatic wavelength conversion in a Silicon photonic ring resonator. This wavelength conversion process is fundamentally linear[17] and is realized by adiabatically shifting the state of the resonator while the photon is confined within it.[11] It previously has been demonstrated using bright classical light in Silicon resonators with free-carrier refractive index modulation through optical[17] and electrical[18] carrier injection. Since the wavelength conversion process is linear there are no photon energy or phase matching conditions, therefore there is no dependence on the lights intensity. In fact using both semi-classical and fully quantum-mechanical arguments, the adiabatic wavelength conversion process has been proven to be one-hundred-percent efficient down to the single photon level.[11,19]

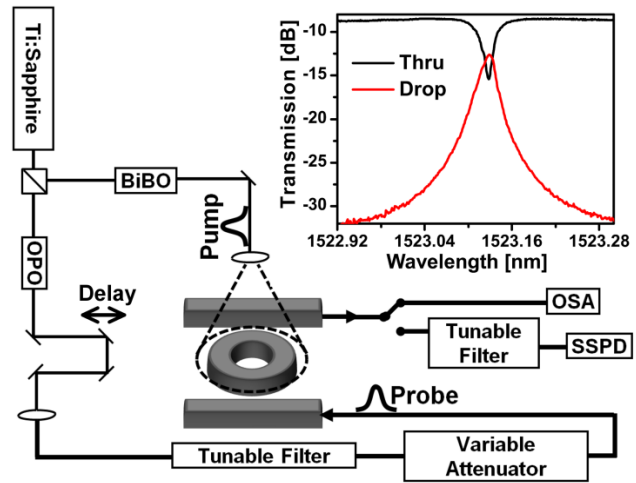

FIG. 1. Schematic of the experimental setup used for single photon adiabatic wavelength conversion. Ti:Sapphire laser (76MHz repetition rate, 100fs pulses, λ=830nm) produces pump pulses (using a BiBO second-harmonic-generation crystal) and probe pulses using an Optical Parametric Oscillator (OPO). A fiber based Fabry-perot tunable etalon (Δλ=0.14nm) filters the probe pulses. The pulses are attenuated to the single photon level using a calibrated variable attenuator. The probe pulses are converted to a new wavelength using the refractive index change induced by the pump that is focused on the ring resonator (inset shows the measured transmission of the ring resonator). The resulting signal is measured either using an Optical Spectrum Analyzer (OSA) or Single Photon Superconducting Detector (SSPD).

In order to demonstrate adiabatic wavelength conversion at the single photon level we use a Silicon ring resonator similar to the one previously used by Preble et al.[17] The measured transmission of one of the rings resonances is seen in the inset of Fig. 1, which shows the transmission past the resonator (Thru) and dropped across the resonator to a second waveguide (Drop). To ensure that the index modulation used to adiabatically shift the resonance is synchronized with the incident photon a

a) Author to whom correspondence should be addressed. Electronic mail: sfpeen@rit.edu

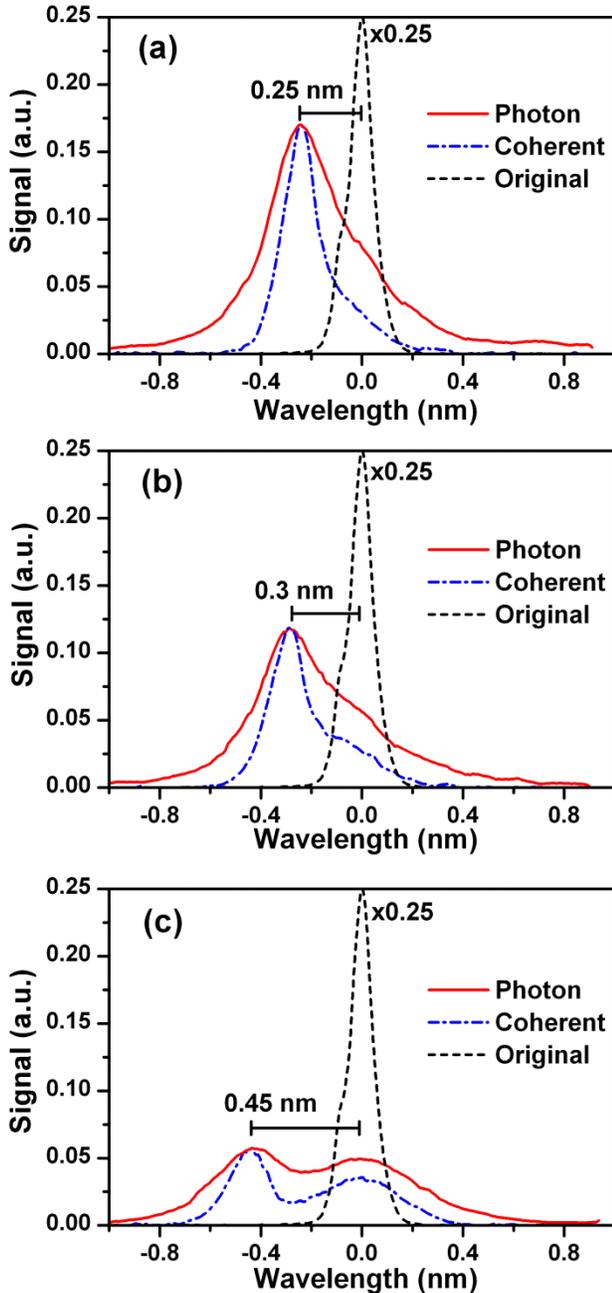

FIG. 2. Adiabatic wavelength conversion at the single photon level and with coherent (bright) light for comparison. The three plots are for wavelength changes of (a) 0.25nm, (b) 0.3nm and (c) 0.45nm. The original probe signal, without wavelength conversion, is shown at ¼ scale (black dashes), The coherent (blue dash-dot) and single photon (red solid) signals are normalized to the original signal. The single photon signal is broader due to the use of a wider bandwidth filter.

100-fs mode-locked Ti:Sapphire laser is used to produce both the pump ($\lambda$=415nm) and single-photon level probe pulses ($\lambda$~1523nm). The pump pulse directly illuminates the top of the ring resonator using a microscope objective, where it is absorbed by the 250nm thick 10µm radius Silicon ring. The absorbed light generates free-carriers which change the refractive index and shifts the resonance according to the well known free-carrier plasma dispersion effect.[17] This index change induces the adiabatic wavelength conversion of the probe pulse. The probe pulse is produced by an Optical Parametric Oscillator (OPO), delayed to match the arrival of the pump pulse, filtered ($\Delta\lambda$=0.14nm) to match the original resonant frequency of the ring resonator and attenuated to a single photon power-level. The average photon number coupled into the ring resonator is approximately $\bar{n} = 0.15$, which minimizes the probability for multi-photon excitations and is a good approximation of a true single photon source. The photons coupled across the ring resonator into the drop port are measured using Superconducting Nanowire Single Photon Detector (SSPD from Scontel, Inc.) with a detection efficiency of $\eta$=5.6% after a second tunable filter ($\Delta\lambda$=0.14nm). The single photon results are directly compared with a bright coherent source by reducing the attenuation and using an Optical Spectrum Analyzer (OSA) with a bandwidth of $\Delta\lambda$=0.05nm to measure the spectrum.

Single photon adiabatic wavelength conversion is observed in Fig. 2 for three different wavelength changes ($\Delta\lambda$=0.25nm, $\Delta\lambda$=0.3nm and $\Delta\lambda$=0.45nm). Each result was obtained by varying the pump pulse power, which is linearly proportional to the wavelength change.[17] We see that in each of the examples that the measured single photon and coherent (bright) signals strongly match. The bandwidth of the single photon signal is broadened because it was filtered using a larger bandwidth filter than used for the coherent signal. In all of the measurements the incident probe pulse is nearly completely shifted from the original wavelength to a new wavelength. This is in agreement with the high conversion efficiency of the adiabatic wavelength conversion process, discussed in more detail below, which fundamentally converts every photon to the new wavelength. However, when using a single resonator there is always a finite probability that some light at the original wavelength will leave the resonator before the wavelength conversion process is initiated by the pump pulse.[17,20] As a result there is a weak signal at the original wavelength, even at the single photon level. This can be minimized by using a resonator with a photon lifetime much longer than the incident pulse or by using a series of resonators.[21]

Noise is an important consideration to single photon wavelength conversion. In contrast to nonlinear wavelength conversion which suffers from competing noise processes, adiabatic wavelength conversion is 100% efficient and as a result is fundamentally noiseless.[11,19] However, in this experiment some of the original wavelength remains due to the incomplete conversion of the probe pulse. This can be regarded as an effective noise but can be spectrally filtered away in practice. It was also observed that there was weak incoherent photon generation by the Silicon waveguide due to the optical pumping process. We measured this emission spectrally and determined that it corresponded to carrier recombination through defects that are formed during the etching process of the Silicon waveguide.[22] However, most of these photons were produced at wavelengths far away from the signals used in the experiment (~1523nm) and were removed using interference filters after the chip. The last source of noise in the experiment is the superconducting single photon detector used.[23] It operated with a relatively low dark count rate of ~1kHz, which is at least one-two orders of magnitude below the detected single photon probe signal at the original wavelength. However, since the overall conversion efficiency is less than 20%, as explained later in this letter, we did subtract dark counts and applied a simple post-measurement averaging of the signals in Fig. 2. In addition, during the measurements each data point comprising the signals was averaged for one-second in order to obtain an accurate measurement of the signal.

An advantage of adiabatic wavelength conversion is that any arbitrary wavelength change can be realized by simply varying the amount of refractive index change in the resonator. Specifically, the wavelength shift and refractive index change are linearly proportional and can be calculated by $\Delta\lambda=(\Delta n/n)\lambda$, where n is the effective index of the waveguide and $\lambda$ is the wavelength of the light.[19] However, in practice the maximum refractive index change is limited by competing absorption processes. Specifically, in this work free-carriers were used which result in a strong absorption effect. This is observed in Fig. 2 where as the wavelength change increases from 0.25nm to 0.45nm the signal at the new wavelength is reduced compared to the original signal. Using the well known free-carrier absorption relations for Silicon we have modeled the effects of absorption as seen in Fig. 3(a).[24] This model is based on the change in the quality factor of the resonator as free-carriers are generated, which changes the relative amount of signal that is absorbed by the carriers as opposed to transmitted to the drop/thru ports. The loaded quality factor of the resonator as fabricated was measured to be Q=61k and by fitting the measured resonant spectrum in the inset of Fig. 1 the intrinsic quality factor was determined to be Q=135k. As seen in Fig. 3(a) free-carrier absorption limits wavelength changes to less than 1nm, at which point more than ~70% of the photons are lost. This effect can be significantly reduced by using a resonator with considerably lower initial quality factor. However, this comes with a tradeoff in the amount of light that escapes the resonator before the conversion process occurs, since a lower quality factor corresponds to a shorter photon lifetime.

In order to further investigate conversion efficiency we measured the conversion spectra as a function of the relative delay between the pump and probe pulse, as seen in Fig. 3(b). Specifically, the number of photons at the original wavelength and at the new wavelength were obtained by integrating the signal at the original (Original $\lambda$) and new (New $\lambda$) wavelengths, as the relative delay between the pump and probe pulses is varied. These integrated signals are normalized to the original signal obtained at the drop port without wavelength conversion (i.e. Fig. 2 – Original curves). In addition, since adiabatic wavelength conversion itself is independent of the exact wavelength change we removed the effect of free-carrier absorption using the model in Fig. 3(a). This allowed us to average the data from multiple wavelength changes in order to obtain a statistical measure of the actual conversion efficiency. We see in Fig. 3(b) that at negative pump-probe delays, which correspond to the probe pulse coming earlier than the pump pulse, the signal at the original wavelength increases. For sufficiently large pump-probe delays (~t<-80ps), greater than the photon lifetime of the cavity ($\tau$~50ps) the signal at the original wavelength approaches a maximum since all of the light leaves the resonator before the conversion process occurs at t=0. Note that the conversion process can be thought of as occurring instantaneously since the pump pulse is very short – 100fs. When the pump-probe delay is near t=-10ps we see that the conversion efficiency to the new wavelength reaches a peak value (η~80%). This is the optimal point where the ring contains the maximum amount of light from the incident probe pulse. As explained earlier and seen in Fig. 2, the ring will never contain 100% of the pulse energy, as some of the optical power from the leading edge of the pulse will have escaped the ring before the power from the trailing edge of the pulse has entered the ring. In practice the exact time when the maximum pulse power will be wavelength converted is a function of the photon lifetime of the resonator and the duration of the input pulse. Here we used a probe pulse with a bandwidth of Δλ=0.14nm, which corresponds to a pulse duration of approximately τ~20ps. This is seen in Fig. 3(b) where the total signal reduces from its maximum value to zero in a time approximately equal to the probe pulse duration. As a result by using later delays it is possible to ensure that a minimal signal at the original wavelength exists, but this comes with a tradeoff in that the light from the probe pulse is not coupled into the resonator before the resonator is shifted to the new wavelength. These inherent tradeoffs dictate the maximum conversion efficiency to approximately 80% in Fig. 3. Including the effects of free-carrier absorption as modeled in Fig. 3(a), and only considering the light that exits the drop port (50% of the total), the measured conversion efficiency for a wavelength change of Δλ=0.15nm is η~27%. We note that this calculation neglects the inherent bandwidth mismatch between the ring resonator and the input pulse, which was solely limited by the available fiber based filters in our setup.

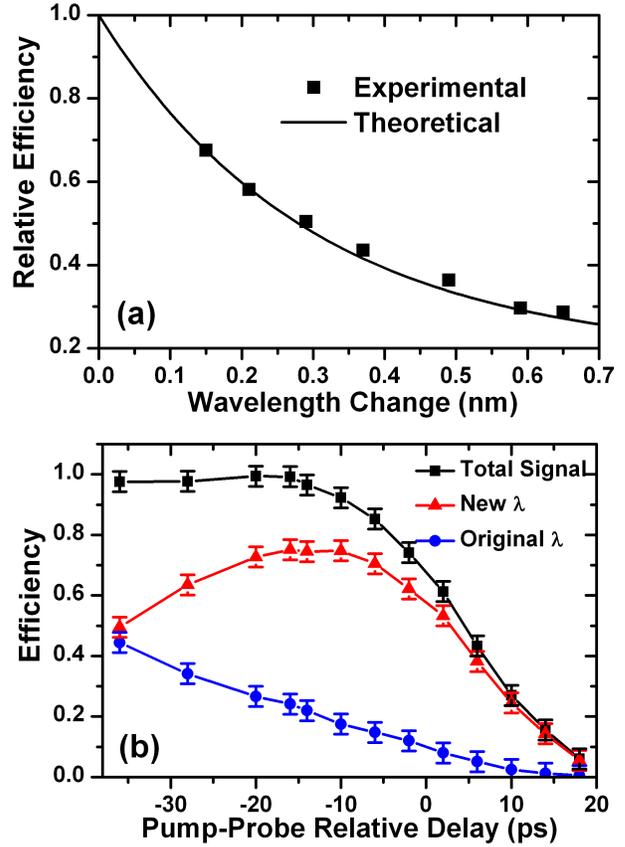

FIG. 3. (a) Relative reduction in conversion efficiencys due to free-carrier absorption vs. wavelength change. (b) Wavelength conversion efficiency as a function of time offset between the pump and the probe, where negative delays correspond to the probe pulse coming earlier than the pump pulse. The data is averaged from multiple wavelength changes. The original $\lambda$ (blue circles) and new $\lambda$ (red triangles) data points are obtained by integrating the spectral signal at the respective wavelengths. The loss due to free carrires, part (a) of the figure, was removed in order to normalize the overall efficiency regardless of wavelength change. The total signal (black squares) is the sum of the New $\lambda$ and Original $\lambda$ data. The lines are provided to guide the eye.

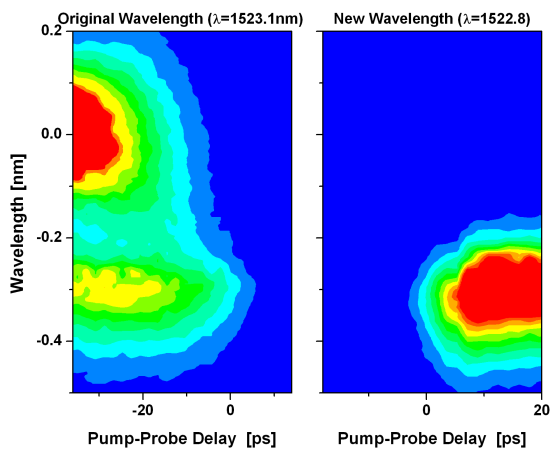

FIG. 4. Measured spectra as the relative delay between the pump and probe pulse is varied (negative delays correspond to the probe coming earlier than the pump). The left panel is when the probe is at the original wavelength of the cavity. The right panel is when the probe is at the new wavelength of the cavity.

A remaining consideration is whether or not the new wavelength is in fact due to adiabatic wavelength conversion or is potentially due to a filtering effect, especially since the probe pulse has a broader bandwidth than the cavity. Fig. 4 shows a delay-wavelength plot for a wavelength change of Δλ=-0.3nm. In the left panel we see that as the pump-probe delay offset increases (negative delays again correspond to the probe pulse coming earlier than the pump) the signal at the original wavelength (λ=0, which corresponds to 1523.1nm in the experiment) is converted to a new wavelength (λ~-0.3nm). And at delays close to t=0 only a signal that the new wavelength exists since most of the photons have been converted. In addition, there is no signal at times after t=0 because this corresponds to the point where the probe pulse arrives *after* the resonator has been shifted to a new resonance by the pump, and consequently cannot couple into the cavity because it is now off-resonance. In order to prove that the new wavelength signal is due to adiabatic wavelength conversion we changed the wavelength of the probe pulse to be on-resonance with the cavities state after the conversion process (λ=1522.8nm). The resulting plot is seen in the right panel of Fig. 4 where there is only a signal at delays t>0, which is the only point where this wavelength can couple into the cavity. Since there is no signal at any other wavelengths or delays we have proven that the new wavelength in the left panel is only due to adiabatic wavelength conversion of the original probe signal.

In summary, we have demonstrated adiabatic wavelength conversion at the single-photon power-level. The process is inherently low noise and very efficient. Any wavelength can be realized by simply tuning the refractive index of the resonator and in this experiment is only limited by the loss due to free-carrier absorption. We note that this loss can be minimized by using a lower quality factor ring resonator. Regardless, the wavelength changes realized here are sufficient for many applications, such as wavelength conversion to adjacent wavelength channels in a wavelength-division-multiplexed quantum network or for tuning a photons wavelength to a quantum dot or atom.[5,6] Lastly, since adiabatic wavelength conversion only requires a refractive index change, other techniques such as cavity opto-mechanics[25] could be used to potentially increase the wavelength change to tens of nanometers.

The authors would like to thank Dr. Gernot Pomrenke, of the Air Force Office of Scientific Research for his support under FA9550-10-1-0217 and NSF under grant ECCS-0824103.